\documentclass[aps,prl,twocolumn,letterpaper,superscriptaddress,10pt,notitlepage]{revtex4-1}
\usepackage{amssymb,amsthm,amsmath,amsfonts}
\usepackage{graphicx,mathptmx}
\usepackage[pdftex,dvipsnames,usenames]{xcolor}
\usepackage[colorlinks=true,urlcolor=blue,citecolor=blue,linkcolor=blue]{hyperref}

\begin{document}

\title{Detector-Agnostic Phase-Space Distributions}

\author{J. Sperling}\email{jan.sperling@uni-paderborn.de}
\affiliation{Integrated Quantum Optics Group, Applied Physics, University of Paderborn, 33098 Paderborn, Germany}

\author{D. S. Phillips}
\affiliation{Clarendon Laboratory, University of Oxford, Parks Road, Oxford OX1 3PU, United Kingdom}

\author{J. F. F Bulmer}
\affiliation{Clarendon Laboratory, University of Oxford, Parks Road, Oxford OX1 3PU, United Kingdom}

\author{G. S. Thekkadath}
\affiliation{Clarendon Laboratory, University of Oxford, Parks Road, Oxford OX1 3PU, United Kingdom}

\author{A. Eckstein}
\affiliation{Clarendon Laboratory, University of Oxford, Parks Road, Oxford OX1 3PU, United Kingdom}

\author{T. A. W. Wolterink}
\affiliation{Clarendon Laboratory, University of Oxford, Parks Road, Oxford OX1 3PU, United Kingdom}

\author{J.~Lugani}
\affiliation{Clarendon Laboratory, University of Oxford, Parks Road, Oxford OX1 3PU, United Kingdom}

\author{S. W. Nam}
\affiliation{National Institute of Standards and Technology, 325 Broadway, Boulder, CO 80305, USA}

\author{A. Lita}
\affiliation{National Institute of Standards and Technology, 325 Broadway, Boulder, CO 80305, USA}

\author{T. Gerrits}
\affiliation{National Institute of Standards and Technology, 325 Broadway, Boulder, CO 80305, USA}

\author{W. Vogel}
\affiliation{Institut f\"ur Physik, Universit\"at Rostock, Albert-Einstein-Stra\ss{}e 23, D-18059 Rostock, Germany}

\author{G. S. Agarwal}
\affiliation{Texas A\&M University, College Station, Texas 77845, USA}

\author{C. Silberhorn}
\affiliation{Integrated Quantum Optics Group, Applied Physics, University of Paderborn, 33098 Paderborn, Germany}

\author{I. A. Walmsley}
\affiliation{Clarendon Laboratory, University of Oxford, Parks Road, Oxford OX1 3PU, United Kingdom}

\date{\today}

\begin{abstract}
	The representation of quantum states via phase-space functions constitutes an intuitive technique to characterize light.
	However, the reconstruction of such distributions is challenging as it demands specific types of detectors and detailed models thereof to account for their particular properties and imperfections.
	To overcome these obstacles, we derive and implement a measurement scheme that enables a reconstruction of phase-space distributions for arbitrary states whose functionality does not depend on the knowledge of the detectors, thus defining the notion of detector-agnostic phase-space distributions.
	Our theory presents a generalization of well-known phase-space quasiprobability distributions, such as the Wigner function.
	We implement our measurement protocol, using state-of-the-art transition-edge sensors without performing a detector characterization.
	Based on our approach, we reveal the characteristic features of heralded single- and two-photon states in phase space and certify their nonclassicality with high statistical significance.
\end{abstract}

\maketitle

\paragraph*{Introduction.|}\hspace{-2ex}%
	The characterization of quantum light is a main challenge one encounters when implementing classically infeasible tasks, such as quantum communication protocols \cite{KMSUZ16,BFV09,RL09}.
	On a more fundamental level, studying the peculiarities of quantized radiation fields leads to a profound understanding of the role of quantum physics in nature in general, and how it is distinct from classical wave theories in particular.
	As in classical systems, quantum-optical phase-space distributions offer a versatile instrument to directly visualize unique features of nonclassical light, such as demonstrated for squeezing \cite{W83,SBRF93,BSM97}.
	Moreover, negativities in certain phase-space functions directly point at quantum properties of light; see, e.g., Refs. \cite{LHABMS01,KVPZB08,LKCGS10,DEWKDKCM13,HSRHMSS16,BFS17}.
	For the above reasons, the representation of quantum light in phase space is one of the most frequently applied methods to characterize nonclassical light.

	However, the estimation of phase-space distributions from experimental data is a cumbersome task.
	Consequently, this reconstruction problem inspired a wide range of research \cite{WVO99,S07,LR09}, leading to sophisticated analytical tools, such as solving inversion problems \cite{T97,SSG09}, employing diverging pattern functions \cite{R96,LMKRR96}, performing maximum-likelihood estimations \cite{H97,L04,KWR04}, and using data pattern recognition \cite{RMH10,MIMSRH13}.
	In addition, each family of detection devices has to be equipped with its own precise model to reliably extract information about phase-space functions \cite{WVO99,S07,LR09}.
	This treatment comprises a comprehensive analysis that assesses (i) how a detector responds to incident light \cite{KK64,FW91}, including, e.g., nonlinear detection responses \cite{JA69,AML11}, and (ii) how the light absorption is influenced by a number of possible imperfections, e.g., efficiencies \cite{PM07,WCLMSPTW09}.
	Moreover, applying these methods can also require universally applicable, yet rather demanding theoretical and experimental techniques in practice, such as performing detector tomography and calibration \cite{K80,LS99,F01,AMP04,FLCEPW09,CLPFSMSEPW09,RFZMGLDFE12,PHMH12,BKSSV17}.

	Despite these challenges, phase-space distributions constitute a highly successful approach to revealing nonclassical properties of light \cite{LHABMS01,KVPZB08,LKCGS10,DEWKDKCM13,HSRHMSS16,BFS17}.
	For example, $s$-parametrized quasiprobabilities \cite{CG69,AW70}, as well as their non-Gaussian generalizations \cite{K66,KV10}, can exhibit negativities that are incompatible with classical light.
	Even if a phase-space function does not exhibit negativities, observable patterns render it possible to identify quantum features, for instance, via the nonnegative Husimi function \cite{H40,LGC18,MS04} or through marginal distributions \cite{A93,PLLSZZZKN17}.
	Because of its success, the concept of phase-space functions has been further extended to other physical scenarios; see Refs. \cite{SW18,SV19}.
	To name a few, atomic ensembles \cite{A81,DAS94,MZHCV15,LMKMIW96} and entanglement \cite{STV12,SV09,SMBBS18} have been successfully characterized using quasiprobability distributions.
	Nevertheless, there remains a dependency on well-defined detection schemes and reconstruction algorithms.

	In this contribution, we circumvent the reconstruction problem by devising a measurement protocol that results in detector-agnostic phase-space (DAPS) distributions, which can be directly estimated, encompass known quasiprobabilities, and apply to arbitrary quantum states of light.
	We demonstrate our scheme with transition-edge sensors (TESs), which have sophisticated physics underlying their operation, and analyze our data without relying on any specific detector models.
	Our DAPS functions reveal nonclassical features expected from our heralded multiphoton states with high statistical significance.
	Moreover, the measurement of vacuum alone enables us to predict the unique structures of DAPS distributions as demonstrated for our experimentally generated states.

\begin{figure*}
	\includegraphics[width=\textwidth]{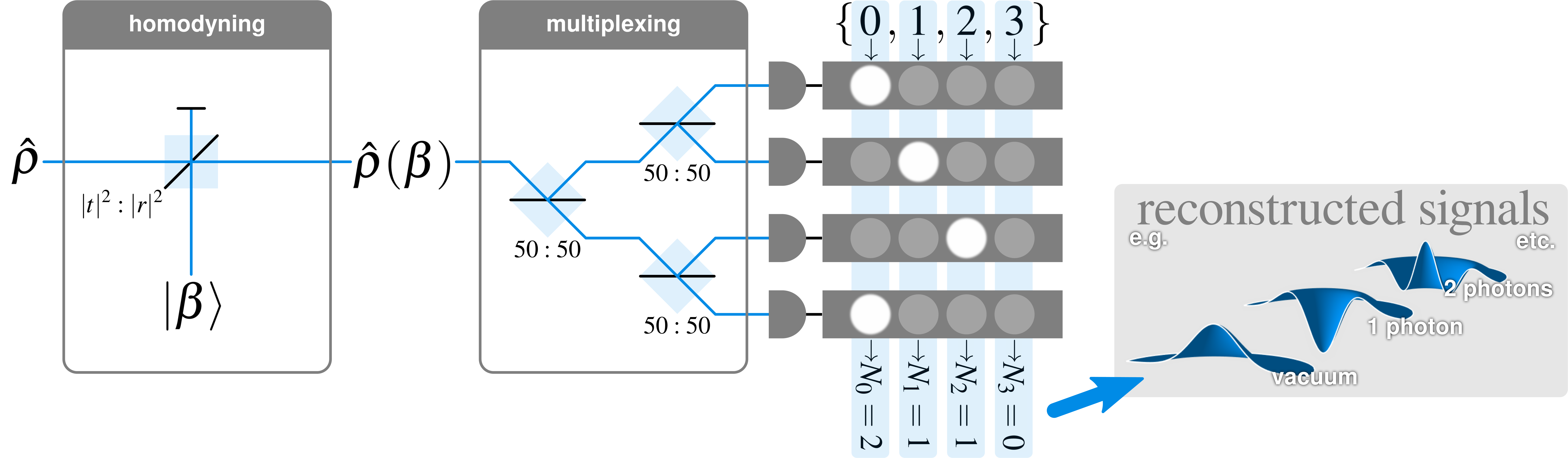}
	\caption{
		Protocol overview.
		A signal state $\hat\rho$ is mixed on a $|t|^2:|r|^2$ beam splitter with a LO $|\beta\rangle$ in an unbalanced homodyning configuration.
		The resulting state $\hat\rho(\beta)$ is fed into a multiplexing scheme (shown for $S=2$ steps).
		Each output beam is measured with a detector that can produce some outcomes (here, $\mathcal K=\{0,\ldots,K=3\}$).
		The resulting statistics $c_{N_0,\ldots,N_K}(\beta)$ is obtained, where $N_k$ counts the number of outcomes $k$ and $N_0+\cdots +N_K=N$.
		From the measured data, we directly estimate our generalized phase-space distributions, cf. the first line in Eq. \eqref{eq:GeneratinFunctions}.
	}\label{fig:Layout}
\end{figure*}

\paragraph*{Theory framework.|}\hspace{-2ex}%
	Our measurement scheme is a combination of unbalanced homodyning \cite{WV96} and a multiplexed detection layout \cite{PTKJ96}; see Fig. \ref{fig:Layout}.
	A signal light field, $\hat\rho$, is mixed with a local oscillator (LO), $|\beta\rangle$, on a beam splitter.
	One of the output states, represented through $\hat\rho(\beta)$, is injected into a multiplexing scheme that consists of $S$ steps.
	In each step, light is split into output fields with the same intensity, which then can be split again.
	The finally obtained $N=2^S$ output beams are individually measured with unknown detectors, which are not specified but assumed to operate in the same manner.
	Each detector returns one of the possible outcomes $\mathcal K=\{0,\ldots,K\}$.
	In Refs. \cite{SCEMRKNLGVAW17,SECMRKNLGWAV17}, we have shown for the multiplexing part that independently of the detector response, the probability to simultaneously measure $N_k$ times the outcome $k$ ($\forall k\in\mathcal K$) follows a quantum version of the multinomial distribution;
	its generalization to $\hat\rho(\beta)$ reads
	\begin{align}
		\label{eq:ClickStatistics}
		c_{N_0,\ldots,N_K}(\beta)=&\left\langle{:}
			\frac{N!}{N_0!\cdots N_K!}\hat\pi_0^{N_0}\cdots\hat\pi_{K}^{N_K}
		{:}\right\rangle_{\hat\rho(\beta)},
	\end{align}
	where ${:}\cdots{:}$ denotes the normal ordering and $\{\hat\pi\}_{k\in\mathcal K}$ is the unknown positive operator-valued measure of the detectors.

	The only assumptions made are a balanced splitting in the multiplexing and identical response functions for the $N$ detectors, including all imperfections.
	We can account for deviations from both assumptions by including a systematic error, directly estimated from asymmetries in the measured data; see the Supplemental Material (SM) for details \cite{supplement}.

	A probability distribution is entirely characterized through its generation function, which can be expressed as
	\begin{align}
	\label{eq:GeneratinFunctions}
	\begin{aligned}
		g_{z_0,\ldots,z_K}(\beta)=&\sum_{N_0,\ldots,N_K}z_0^{N_0}\cdots z_K^{N_K}c_{N_0,\ldots,N_K}(\beta)
		\\
		=&\left\langle{:}
			\left(z_0\hat\pi_0+\cdots+z_K\hat\pi_K\right)^N
		{:}\right\rangle_{\hat\rho(\beta)},
	\end{aligned}
	\end{align}
	for $z_0,\ldots,z_K\in\mathbb R$.
	The second line is a result of the multinomial form of the statistics in Eq. \eqref{eq:ClickStatistics}.
	One salient feature is that classical light fields have a nonnegative generation function $g_{z_0,\ldots,z_K}$.
	To see this, first recall that a classical light field is described through a nonnegative Glauber-Sudarshan distribution \cite{G63,S63}, which is not affected by displacements and describes a state as a statistical mixture of coherent states.
	Furthermore, for all even $N$, we can define the operator $\hat f=\hat f^\dag=\left(z_0\hat\pi_0+\cdots+z_K\hat\pi_K\right)^{N/2}$.
	Since for any nonnegative Glauber-Sudarshan function $\langle {:}\hat f^\dag\hat f{:}\rangle\geq0$ holds true \cite{TG65,M86,VW06,A12}, we conclude
	\begin{align}
		\label{eq:Negativity}
		g_{z_0,\ldots,z_K}(\beta)\stackrel{\mathrm{cl.}}{\geq}0.
	\end{align}
	A violation of this inequality certifies the nonclassicality of the signal light, $\hat \rho$.
	We can also define a special case of this generating function,
	\begin{align}
		\label{eq:GeneratinFunctionSpecial}
		G_z(\beta)=&g_{1,z,z^2,\ldots,z^K}(\beta).
	\end{align}
	Similarly to the expression in Eq. \eqref{eq:GeneratinFunctions}, $G_z$ is straightforwardly estimated from the measured detector outcomes $c_{N_0,\ldots,N_K}(\beta)$ by setting $z_k=z^k$, and $G_z$ is nonnegative for classical light.

	As an example, we may consider photocounting \cite{KK64}.
	Although this model is not required for our approach and does not apply to our experiment (TESs have a finite photon-number resolution, a non-unit detection efficiency, and a nonlinear response function \cite{SECMRKNLGWAV17}), it demonstrates how $G_z$ generalizes the concept of well-known phase-space distributions.
	For photocounting, we find \cite{supplement,WM64}
	\begin{align}
		\label{eq:Pfct}
		G_z(\beta)=&\langle{:}e^{
			-[1-z]\eta\hat n
		}{:}\rangle_{\hat\rho(\beta)}
		=\frac{\pi(1-s)}{2}P\left(\frac{r}{t}\beta;s\right),
	\end{align}
	with $\eta$ and $\hat n$ being the efficiency and the photon-number operator, respectively, and $s=1-2/[\eta|t|^2(1-z)]$ \cite{commentsParameter}.
	Thus, $G_z(\beta)$ resembles the $s$-parametrized distributions $P(r\beta/t;s)$ \cite{CG69,AW70}.
	Beyond photoelectric detectors, we refer to $g_{z_0,\ldots,z_K}$ and $G_z$ as DAPS distributions as Eqs. \eqref{eq:GeneratinFunctions} and \eqref{eq:GeneratinFunctionSpecial} apply without any knowledge of the measurement operators $\{\hat\pi_k\}_{k\in\mathcal K}$.
	In this context, it is worth emphasizing that the first line in Eq. \eqref{eq:GeneratinFunctions} enables the estimation of our DAPS distributions as a result of the measured coincidence statistics $c_{N_0,\ldots,N_K}(\beta)$ alone.

\paragraph*{Implementation.|}\hspace{-2ex}%
	By implementing a single multiplexing step, $N = 2^S= 2$ for $S=1$, we demonstrate how to apply our theoretical framework of DAPS distributions.
	To realize our protocol in Fig. \ref{fig:Layout}, we produce heralded photon states $\hat\rho$ and different LO amplitudes $\beta$.
	The detectors used for the multiplexing measurement and the heralding are TESs, which count photons up to a maximal number $K$.
	In the following, we describe the experimental setup \cite{supplement}.

	Femtosecond pulses with a $100\,\textrm{kHz}$ repetition rate from a titanium sapphire laser are coupled into two separate, periodically poled potassium titanyl phosphate (ppKTP) waveguides.
	Both pulses are filtered to a full-width at half-maximum of $\pm2\,\textrm{nm}$ using angle-tuned bandpass filters.
	With the first ppKTP waveguide, prepare the signal $\hat\rho$.
	When filtering the pump at $775\,\textrm{nm}$ the waveguide produces two-mode squeezed vacuum in approximately a single spatio-temporal mode via type-II parametric down conversion (PDC) \cite{ECMS11}.
	The signal mode at $1554\,\textrm{nm}$ and the herald mode at $1547\,\textrm{nm}$ are separated with a polarizing beam splitter, then filtered and coupled into single-mode optical fibers.
	The herald mode is then sent to a single TES detector.
	With the second ppKTP waveguide, we prepare the LO.
	In contrast to the signal state generation, we filter the pump at $783\,\textrm{nm}$ and stimulate the PDC process by seeding it with $2\,\mathrm{ns}$ pulses carved with an electro-optic modulator from a $1580\,\textrm{nm}$ continuous-wave laser.
	Because of the strong seed signal, this nonlinear mixing generates coherent light to an excellent approximation in the polarization mode orthogonal to the seed \cite{LS13}.
	We separate the LO from the seed with a polarizing beam splitter, then pass through a bandpass filter at $1554\,\textrm{nm}$.
	By pumping the two waveguides at different wavelengths, we are able to create an LO that is well mode-matched to the signal using a seed laser that is detuned from the heralding mode.
	This avoids a potential source of noise due to the seed laser passing through the filters for the heralding TES.
	The generated LO is attenuated to the single-photon level and coupled into single-mode optical fiber.
	Crucially, this process prepares an LO with Poissonian photon statistics with a measured second-order correlation function $g^{(2)}(0)$ of $1.005\pm0.002$.

	Finally, the LO $|\beta\rangle$ and signal $\hat\rho$ are combined on a $90:10$ fiber beam splitter.
	We consider the port that uses $|r|^2=10\%$ of the LO and transmits $|t|^2=90\%$ of the signal.
	The light from this port, $\hat\rho(\beta)$, is then impinged on a $50:50$ fiber beam splitter for realizing a multiplexing step;
	both outputs are then sent to two separate TESs.
	See Fig. \ref{fig:Layout}.
	
	Our experiment uses three TES detectors that can have efficiencies above $\eta=90\%$ \cite{LMN08}.
	TESs are superconducting photon-number-resolving detectors that we operate in a dilution refrigerator at a temperature of around $80\,\mathrm{mK}$.
	Their response is amplified using an array of superconducting quantum interference devices \cite{WM91}, followed by further amplification and filtering at room temperature.
	This electrical signal is read by an analogue-to-digital converter and processed using a matched filter technique \cite{FCMPSW00}, which outputs a single value when triggered by a clock signal from the laser.
	We bin these values to assign a photon number.
	It should be noted that it is possible to extract slightly more accurate estimates of photon number, however, using more sophisticated signal processing techniques, yet without affecting the applicability of the DAPS distribution approach \cite{SECMRKNLGWAV17,HMGHLNNDKW15}.

	We record the binned outcome at all three TESs for various LO amplitudes ($|\beta|^2$ from $0$ to ${\sim}28$ in steps of ${\sim}1$).
	The amplitude is controlled by varying the seed laser power.
	To obtain data for a specific heralded state $\hat\rho$, we consider the subset of trials with the appropriate detection outcome (i.e., heralding bin $k_h$) at the herald TES.

\paragraph*{Verification of nonclassicality.|}\hspace{-2ex}%
	In a first step, we apply our DAPS distribution to uncover nonclassical features of our prepared states through the violation of condition \eqref{eq:Negativity}.
	The optimal negativity we obtain from the DAPS function [Eq. \eqref{eq:GeneratinFunctions}] is given by the minimum
	\begin{align}
		\label{eq:MinimalPhaseSpace}
		g_{\min}=\min_{\beta}\min_{\substack{z_0,\ldots,z_K:\\|z_0|^2+\cdots+|z_K|^2\leq1}} g_{z_0,\ldots,z_K}(\beta).
	\end{align}
	To assess the quality of this approach, we compared our verification of nonclassicality with other methods.
	In Ref. \cite{SCEMRKNLGVAW17}, we demonstrated that a correlation matrix, $M$, obtained from the measured statistics in Eq. \eqref{eq:ClickStatistics}, is positive semidefinite for classical light, described through a nonnegative minimal eigenvalue $\mu_{\min}$ of $M$.
	The resulting notion of sub-multinomial light, $\mu_{\min}<0$, was shown to be a better figure of merit than other means of verifying nonclassicality \cite{SECMRKNLGWAV17}, such as sub-Poisson light \cite{M79,SM83} and sub-binomial light \cite{SVA12,BDJDBW13}.

	The comparison of $g_{\min}$ and $\mu_{\min}$ for our data is shown in Table \ref{Tab:SomeFigures} for different heralding bins $k_h$.
	For the heralded one-photon (two-photon) states, we confirm $g_{\min}<0$ with $9$ ($6$) standard deviations, while the sub-multinomial behavior is less significant, $5$ ($3$) standard deviations.
	For the vacuum state, i.e., $k_h=0$, both measures are consistent with the classical expectation, $g_{\min}=0=\mu_{\min}$.

\begin{table}[t]
	\caption{
		For different heralding outcomes, $k_h$, we show the nonclassicality criteria $\mu_{\min}<0$ and $g_{\min}<0$.
		$g_{\min}$ is defined in Eq. \eqref{eq:MinimalPhaseSpace}.
		$\mu_\mathrm{min}$ is the minimal eigenvalue to the second-order correlation matrix $M$ defined in Eq. (6) of Ref. \cite{SCEMRKNLGVAW17}; see also the SM \cite{supplement}.
		``$-0$'' indicates a slightly negative mean value that rounds to zero.
	}\label{Tab:SomeFigures}
	\begin{tabular}{p{.2\columnwidth}p{.35\columnwidth}p{.35\columnwidth}}
		\hline\hline
		$k_h$ & $\mu_{\min}$ & $g_{\min}$
		\\\hline
		0& \hspace{2ex}$(-0 \pm 9)\times 10^{-4}$ & \hspace{3ex}$(-0 \pm 2)\times 10^{-9}$
		\\
		1& $-0.15 \pm 0.03$ & $-0.026 \pm 0.003$
		\\
		2& $-0.10 \pm 0.03$ & $-0.017 \pm 0.003$
		\\\hline\hline
	\end{tabular}
\end{table}

\begin{figure*}
	\includegraphics[width=\textwidth]{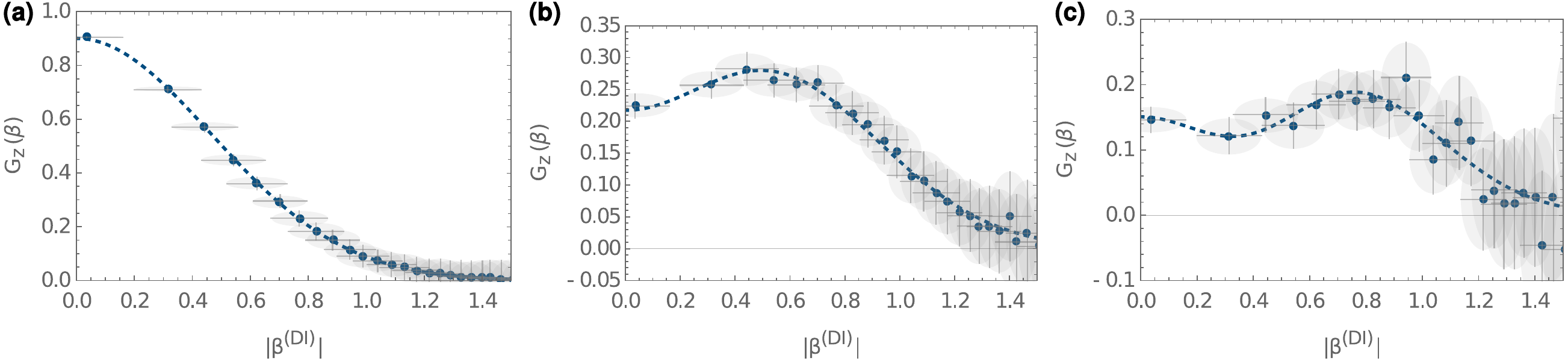}
	\caption{
		Reconstructed DAPS distributions $G_z(\beta)$ [Eq. \eqref{eq:GeneratinFunctionSpecial}] as a function of $|\beta^\mathrm{(DI)}|$ [Eq. \eqref{eq:DefinitionAmplitude}].
		We choose $z=-1.5$ as it would correspond to a Wigner function ($s=0$) in the case of photocounting under the assumption of almost no loss, $\eta\approx 90\%$.
		From left to right, (a)--(c), heralded $k_h=0,1,2$-photon states are shown.
		The dashed lines show the fit to a model inferred from the data obtained by blocking the signal [Eq. \eqref{eq:VaccumConvolution}].
		The defining structures of the heralded (b) single-photon and (c) two-photon states are the oscillating patterns near the origin $|\beta^\mathrm{(DI)}|=0$.
	}\label{fig:NewFig}
\end{figure*}

\paragraph*{Reconstructed distributions.|}\hspace{-2ex}%
	From the data, we can directly estimate our DAPS distributions.
	The results of our extended analysis are shown in Fig. \ref{fig:NewFig}.
	The estimation procedure is this:
	we run the experiment twice, once with the signal blocked and once with the signal unblocked.

	To have full detector-agnostic approach, we first define a detector-independent coherent amplitude,
	\begin{align}
		\label{eq:DefinitionAmplitude}
		|\beta^\mathrm{(DI)}|=\sqrt{\sum_{N_0,\ldots,N_K}[0N_0+\cdots +KN_K]c^\mathrm{(vac)}_{N_0,\ldots,N_K}(\beta)},
	\end{align}
	which is given by the statistics $c^\mathrm{(vac)}_{N_0,\ldots,N_K}(\beta)$ measured by blocking the signal \cite{supplement}.
	In case of photocounting, this gives $|\beta^\mathrm{(DI)}|=\sqrt\eta|r||\beta|$.
	As we do not record a phase, we consider full phase randomization.
	This does not affect the DAPS distributions of our heralded photon states.
	In Fig. \ref{fig:NewFig}, our DAPS distributions $G_z$ are shown as a function of the amplitude in Eq. \eqref{eq:DefinitionAmplitude}, determined by means of the vacuum measurement.

	The same measurement renders it possible to theoretically predict the DAPS distribution of arbitrary states.
	Namely, a general DAPS distribution can be described as a convolution of the measured vacuum distribution $G_z^\mathrm{(vac)}$ and the Glauber-Sudarshan distribution $P(\beta';1)$ of the state under study \cite{supplement},
	\begin{align}
		\label{eq:VaccumConvolution}
		G_z(\beta)=\int d^2\beta'\, P(\beta';1) G^\mathrm{(vac)}_{z}\left(\beta-\frac{t}{r}\beta'\right).
	\end{align}
	In our case, the Gaussian shape of $G_z^\mathrm{(vac)}$ implies that heralded single-photon (two-photon) states should follow a Gaussian distribution multiplied with a first-order (second-order) polynomial in $|\beta^\mathrm{(DI)}|^2$.
	In Fig. \ref{fig:NewFig}, this prediction (dashed lines) is confirmed as it correctly represents the DAPS distributions of the measured heralded photon states.
	The heralding to $k_h=1$ gives a characteristic dip at the origin $|\beta^\mathrm{(DI)}|=0$, and the two-photon case, $k_h=2$, leads to additional oscillations together with the appearance of a peak at the origin.
	We emphasize that the functional behavior $\beta\mapsto G_z(\beta)$ depends on the measurement operators, but the estimation of $G_z(\beta)$ is done without any specification of the detector operators, according to the first line in Eq. \eqref{eq:GeneratinFunctions}.
	Moreover, we are able to characterize defining features of other states without any other prior knowledge about the detectors from the data obtained using the vacuum state input [Eq. \eqref{eq:VaccumConvolution}].

	Based on our reconstruction, we were able to determine a number of other properties of the experimentally produced states \cite{supplement}.
	For instance, we can determine how well the DAPS distributions enable us to perform a quantum state discrimination task.
	The single- and two-photon states [plots (b) and (c) in Fig. \ref{fig:NewFig}] can be distinguished from each other with more than $98\%$ certainty.
	Furthermore, we found that, for $z<-2.4$, the central dip of $G_z$ becomes negative, similar to the behavior of other phase-space quasiprobabilities.
	The negativity has the highest statistical significance for $z=-4.85$, where $G_{z}(0)=-0.51\pm0.08$ is more than 6 standard deviations below the classical threshold of zero.

\paragraph*{Summary and discussion.|}\hspace{-2ex}%
	We have developed a theory and realized an experiment to characterize quantum light by means of phase space that functions for any type of detector and without performing a prior detector characterization.
	Our framework is based on the generating function derived from the properties of a balanced linear optical network, enabling our DAPS distribution, to be directly estimated from measured correlations and that applies to arbitrary states.

	To demonstrate this concept, we showed that a single multiplexing step is already sufficient for applying our method.
	This renders it possible to verify the nonclassicality of multiphoton states based on DAPS distributions, which results in greater statistical significance than obtained with earlier approaches \cite{SCEMRKNLGVAW17,SECMRKNLGWAV17}, which themselves already outperformed previous quantifiers of nonclassicality.
	Moreover, our approach encompasses prominent phase-space quasiprobabilities and straightforwardly generalizes to multimode light.

	Our general theory also includes more recent phase-space functions based on on-off detectors \cite{LSV15}, constituting the special case $K=1$ and being applicable to off-the-shelf detectors (e.g., avalanche photodiodes in Geiger mode and single-photon nanowire detectors);
	to prove this, see the corresponding experiment with $S=1,2,3$ multiplexing steps \cite{BTBSSV18}.
	Furthermore, recent advancements in detector technology (see, e.g., Refs. \cite{DetNew1,DetNew2,DetNew3,DetNew4}) offer new photon counters to which our detector-agnostic framework is also readily applicable.

	With our approach, we are further able to predict defining phase-space features of any states by measuring vacuum as a reference.
	Thus, we can compare a target state with the actually reconstructed DAPS distribution, thus enabling us to estimate other quantum properties as well.
	As a practical example, a state discrimination task based on our DAPS distributions resulted in distinguishing one- and two-photon states with almost unit certainty, despite high losses in our setup.
	It is also worth noting that our DAPS distribution includes the full quantum information of the state that is accessible with the detectors used and does not require demanding reconstruction algorithms and detection models.

	Our experiment comprises state-of-the-art detectors combined with an advantageous method to create coherent states, well mode matched to our signal.
	As our method is detector agnostic, the detector efficiency need not be specified or even known;
	the number of data points merely has to be sufficient to produce statistically meaningful results.
	Also, our approach is not restricted to any specific states;
	currently, we are mainly limited by the available sources of nonclassical light.

	In the future, recording the LO's phase would be beneficial for applying our scheme to phase-sensitive nonclassical states as well.
	Furthermore, generalizing other interferometric measurement schemes in a detector-agnostic manner is feasible, e.g., as done for on-off detectors \cite{SVA15}.
	In addition, we encounter the imperfections stemming from imbalances by assigning systematic errors.
	It may be possible to avoid this by using more sophisticated strategies \cite{LFPR16}.

	Generalized phase-space distributions are becoming increasingly important in identifying vastly different notions of quantumness; see Refs. \cite{SW18,SV19} for thorough overviews.
	To date, however, such universally applicable techniques are also highly dependent on the particular response of the detectors.
	Our DAPS approach, however, sets a precedence that such limitations can be overcome in theory and experiment.

	In conclusion, our detector-agnostic framework provides a universally applicable approach to the robust characterization of quantum light in phase space under conditions where detailed knowledge of the measurement apparatus is not available, and forms a basis for future research.

\paragraph*{Acknowledgments.|}\hspace{-2ex}%
	The authors are grateful to William R. Clements for helpful discussions and Jelmer J. Renema  for his assistance with the installation of the cryogenic infrastructure.
	The authors also thank Scott Glancy, Arik Avagyan, and Tim Bartley for valuable comments.
	The Integrated Quantum Optics group acknowledges financial support from the Gott\-fried Wilhelm Leibniz-Preis (Grant No. SI1115/3-1).
	This work received funding through the Networked Quantum Information Technologies (NQIT) hub (part of the UK National Quantum Technologies Programme) under Grant No. EP/N509711/1.
	G. S. T. acknowledges financial support from the Natural Sciences and Engineering Research Council of Canada and the Oxford Basil Reeve Graduate Scholarship.
	A. E. is supported by EPSRC (project EP/K034480/1 BLOQS).
	T. A. W. W. is supported by Fondation Wiener - Anspach.
	J. L. thanks the European Commission (H2020-FETPROACT-2014 grant QUCHIP).
	I. A. W. acknowledges ERC (Advanced Grant MOQUACINO).
	This work was supported by the Quantum Information Science Initiative (QISI).


\onecolumngrid
\section*{Supplemental Material}\label{sec:Appendices}
\appendix
\twocolumngrid

\section{Details on the experiment}\label{sec:Experiment}

\begin{figure}[t]
	\includegraphics[width=\columnwidth]{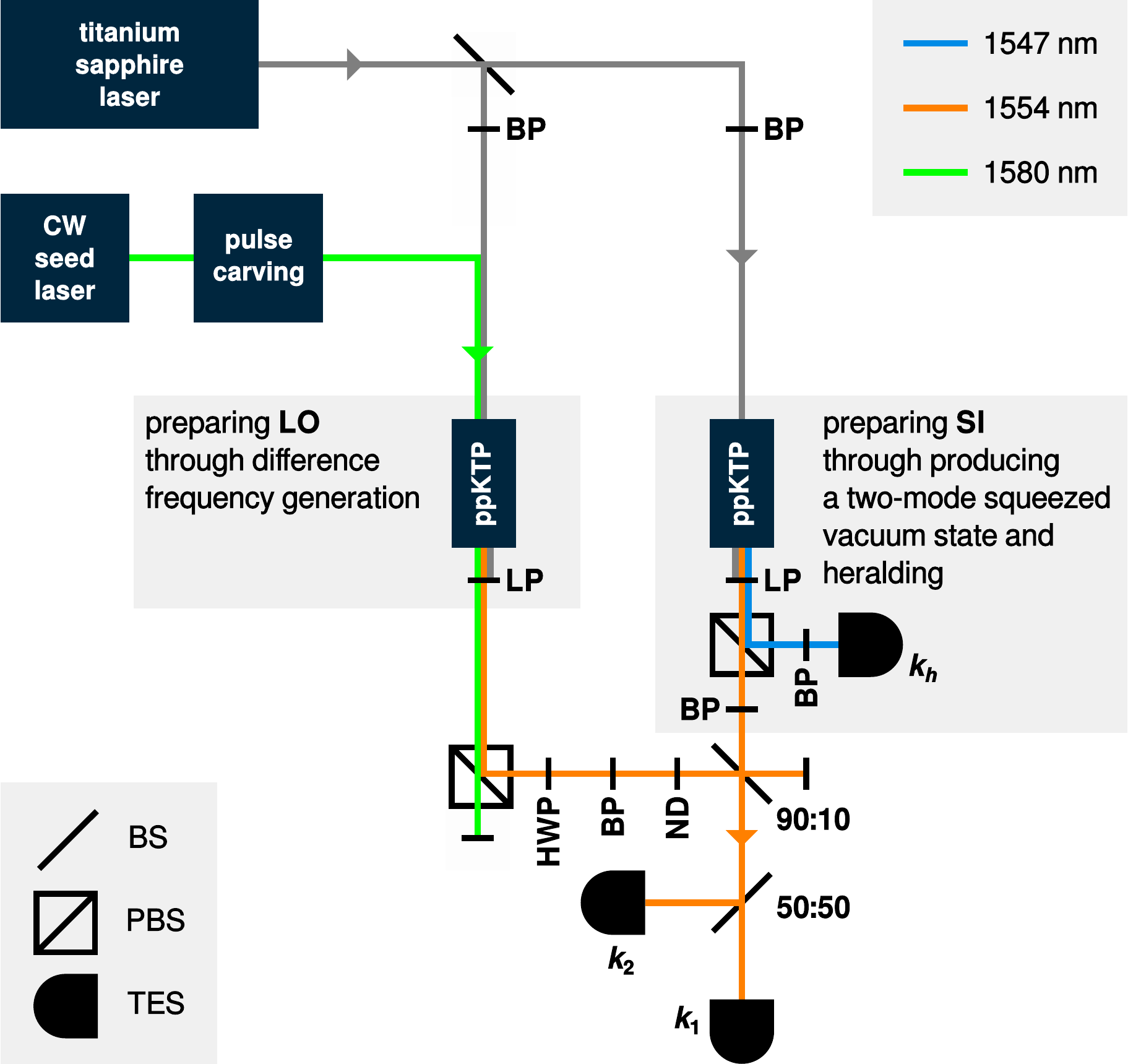}
	\caption{
		Outline of the setup;
		see Sec. \ref{sec:Experiment} for the full description.
		BP: bandpass filter, BS: beam splitter, CW: continuous-wave, HWP: half-wave plate, ppKTP: periodically poled potassium titanyl phosphate, LP: longpass filter, ND: neutral-density filter, PBS: polarizing beam splitter, TES: transition-edge sensor.
	}\label{fig:Setup}
\end{figure}

	The experimental setup is shown in Fig. \ref{fig:Setup}.
	Our pump laser is a titanium sapphire (Ti:Saph) regenerative amplifier that generates femtosecond pulses (center wavelength $780\,\mathrm{nm}$, full width at half maximum [FWHM] $15\,\mathrm{nm}$) at a rate of $100\,\mathrm{kHz}$.
	This rate is chosen to accommodate the thermal relaxation time (${\sim}10\,\mathrm{\mu s}$) of the transition-edge sensors (TESs).
	We split the pulses into two paths, each pumping a periodically poled potassium titanyl phosphate (ppKTP) waveguide.

	We pump the first waveguide (right in Fig. \ref{fig:Setup}) using filtered (center $775\,\mathrm{nm}$, FWHM $2\,\mathrm{nm}$) pulses from the Ti:Saph.
	Pumping the ppKTP waveguide generates two-mode squeezed vacuum via type-II parametric down-conversion.
	The pump is then discarded using a longpass filter.
	The two down-converted modes (signal $1554\,\mathrm{nm}$, idler $1547\,\mathrm{nm}$) are orthogonally polarized and separated by a polarisation beam splitter.
	Each mode is sent through a bandpass filter (FWHM $10\,\mathrm{nm}$).
	The idler mode is sent to a TES detector to herald photon-number states in the signal mode (heralding efficiency ${\sim}40\%$) by postselecting to a specific outcome $k_h$.

	In the second waveguide (left in Fig. \ref{fig:Setup}), we prepare the local oscillator (LO).
	Since the first and second waveguides have slightly different phase-matching properties, a different pump spectrum (center $783\,\mathrm{nm}$, FWHM $2\,\mathrm{nm}$) is used.
	This pump spectrum is chosen to maximize the spectral overlap between the LO and signal (SI).
	We also carve $2\,\mathrm{ns}$ square seed pulses from a continuous-wave laser (center $1580\,\mathrm{nm}$) using an electro-optic modulator.
	The pump and seed pulses are temporally overlapped and coupled into the second ppKTP waveguide.
	Through difference frequency generation, the LO is generated in the polarization orthogonal to the seed.
	The LO is separated from the seed using a polarisation beam splitter.
	As before, we discard the pump by a longpass filter.
	The LO's polarization is adjusted with a half-wave plate to match the SI's polarization.
	Then, the LO is sent through a bandpass filter (FWHM $10\,\mathrm{nm}$) to further eliminate seed light as well as increase the spectral overlap with the signal.
	Finally, neutral-density filters attenuate the LO to the single-photon level.

	The SI and LO are combined on a 90:10 beam splitter.
	The resulting light field of one output then enters the multiplexing step (50:50 beam splitter) and the then resulting beams are measured with two TESs.
	The recorded coincidences give the detection events $E(k_1,k_2)$, which we use for our analysis.

	In addition, we characterized the mode overlap of the signal and LO by combining the two on a 50:50 beam splitter.
	We consider the specific case of a single photon ($k_h=1$) and a weak LO ($|\beta|\ll 1$).
	By scanning the delay between the SI and LO, we expect to measure a Hong-Ou-Mandel-type dip in two-fold coincidences at the output of the beam splitter.
	We measured a dip of ${\sim}80\%$ visibility, suggesting that the mode overlap is at least $80\%$.
	By blocking the signal, this setup constitutes a Hanbury Brown-Twiss interferometer that allows us to measure the LO's $g^{(2)}(0)$.
	We measured $g^{(2)}(0) = 1.005 \pm 0.002$, which is consistent with the expected Poisson distribution for the LO's photon statistics.

\section{Details on the theory}\label{sec:Theory}

\subsection{General approach}

	Let us formulate some additional details on the theory.
	As we can expand any SI state in the Glauber-Sudarshan decomposition, $\hat\rho=\int d^2\alpha\,P(\alpha)|\alpha\rangle\langle \alpha|$, it is sufficient to consider the propagation of coherent states $|\alpha\rangle$.
	Our detection scheme consists of a combination of the SI with the LO state $|\beta\rangle$ on a beam splitter, the multiplexing, and the detection.

	Applying a beam splitter transformation, we map an input, consisting of SI and LO, as follows: $|\alpha\rangle\otimes|\beta\rangle\mapsto |t\alpha-r\beta\rangle\otimes|r^\ast\alpha+t^\ast\beta\rangle$, where $t$ and $r$ define the trasmissivity and reflectivity ($|t|^2+|r|^2=1$).
	When tracing over the second mode, we obtain the state that enters the multiplexing stage,
	\begin{align}
		\label{eq:DisplacedState}
		\hat\rho(\beta)=\int d^2\alpha\, P(\alpha)|t\alpha-r\beta\rangle\langle t\alpha-r\beta|.
	\end{align}

	Further, the multiplexing distributes the coherent state components in Eq. \eqref{eq:DisplacedState} among the $N=2^S$ output beams, where $S$ is the depth of the multiplexing scheme, $|\gamma\rangle\mapsto |\gamma/\sqrt N\rangle^{\otimes N}$, resulting in $\int d^2\alpha\, P(\alpha)[|(t\alpha-r\beta)/\sqrt N\rangle\langle (t\alpha-r\beta)/\sqrt N|]^{\otimes N}$.
	Using some combinatorics (see Ref. \cite{SECMRKNLGWAV17} for details), we find
	\begin{align}
	\begin{aligned}
		c_{N_0,\ldots,N_K}(\beta)=&\int d^2\alpha\,P(\alpha)\frac{N!}{N_0!\cdots N_K!}
		\\&\times
		\prod_{k=0}^K{\underbrace{\left\langle \frac{t\alpha-r\beta}{\sqrt N}\right|\hat\pi_k\left|\frac{t\alpha-r\beta}{\sqrt N}\right\rangle}_{\stackrel{\mathrm{def.}}{=}p_k\left(\frac{t\alpha-r\beta}{\sqrt N}\right)}}^{N_k},
	\end{aligned}
	\end{align}
	where $\{\hat\pi_k\}_{k=0,\ldots,K}$ is an unknown positive operator-valued measure (POVM) that describes the detector.

	Consequently, the two types of generating functions under consideration read
	\begin{align}
		\label{eq:GeneratingFunction}
		g_{z_0,\ldots,z_K}(\beta)=&\int d^2\alpha\,P(\alpha)\left[
			\sum_{k=0}^K z_kp_k\left(\frac{t\alpha-r\beta}{\sqrt N}\right)
		\right]^N
	\intertext{and}
		\label{eq:GeneratingFunction2}
		G_z(\beta)=&\int d^2\alpha\,P(\alpha)\left[
			\sum_{k=0}^K z^kp_k\left(\frac{t\alpha-r\beta}{\sqrt N}\right)
		\right]^N,
	\end{align}
	using the Glauber-Sudarshan $P$ function.
	As long as $N$ is even and $P\geq0$, both expressions are necessarily nonnegative.

\subsection{Photoelectric counting with loss}

	Let us analyze our scheme for the special case of photoelectric counting.
	A simple photoelectric detection is described through POVM elements $\hat\pi_k={:}e^{-\eta\hat n}(\eta\hat n)^k/k!{:}$ for $k=0,1,\ldots$ ($K=\infty$), where $\eta$ is the quantum efficiency and $\hat n$ is the photon-number operator.
	In this scenario, the generating function in Eq. \eqref{eq:GeneratingFunction2} can be further evaluated \cite{WM64} and reads
	\begin{align}
	\label{eq:Photoelectric}
	\begin{aligned}
		G_z(\beta)=&\int d^2\alpha\,P(\alpha)\exp\left(
			-[1-z]|t|^2\eta\left|\alpha-\frac{r}{t}\beta\right|^2
		\right)
		\\
		=&\left\langle{:}
			\exp\left[
				-(1-z)|t|^2\eta\hat n\left(\frac{r}{t}\beta\right)
			\right]
		{:}\right\rangle_{\hat\rho},
	\end{aligned}
	\end{align}
	where $\hat n(\gamma)$ is the displaced photon-number operator.
	Since we have $P(\gamma;s)=2(\pi[1-s])^{-1}\langle{:}e^{-2\hat n(\gamma)/[1-s]}{:}\rangle_{\hat\rho}$ \cite{WV96}, the above expression can be related to $s$-paramterized distributions.

	Also note that according to the characterization performed in Ref. \cite{SECMRKNLGWAV17} (Sec. III), our TESs are more precisely described through POVMs of the form $\hat\pi_k={:}e^{-\Gamma(\hat n)}\Gamma(\hat n)^k/k!{:}$ for $k=0,\ldots, K-1$ and $\hat\pi_K=\hat 1-\sum_{k=0}^{K-1}\hat\pi_k$.
	Therein, $K<\infty$ reflects the finite photon-number resolution, and the response function $\Gamma$ has a nonlinear form, $\Gamma(\hat n)\approx\eta\hat n+\eta^{(2)}\hat n^2$, where the quantum efficiency $\eta$ is not one ($\eta<1$) and the nonlinear contribution does not vanish ($\eta^{(2)}\neq0$).
	As $\eta^{(2)}$ is small, the nonlinear behavior mainly affects higher LO and SI intensities.

\section{Coincidences and systematic errors}\label{sec:Coincidence}

	A single multiplexing step, $N=2$, was implemented.
	Thus, it is convenient to formulate the data processing in terms of measured coincidences.
	For this purpose, we denote with $E(k_1,k_2)$ the number of coincidence events for the measurement outcomes $k_1$ and $k_2$ ($k_1,k_2\in\{0,\ldots,K\}$), resembling the detection bins of the TESs $1$ and  $2$, respectively.
	$E=\sum_{k_1,k_2}E(k_1,k_2)$ defines the total number of events.

	The coincidences are directly related to the desired quantum version of a multinomial distribution, $c_{N_0,\ldots,N_K}$, cf. Eq. (1) in the main text.
	Since $N_0+\cdots+N_K=N=2$, we have
	\begin{align}
		c_{N_0,\ldots,N_K}
		=\left\{\begin{array}{ll}
			c_{0,\ldots,0,N_i=2,0,\ldots,0} & \text{ for }0\leq i\leq K,
			\\
			c_{0,\ldots,0,N_i=1,0,\ldots,0,N_j=1,0,\ldots,0} & \text{ for }0\leq i<j\leq K,
			\\
			0 & \text{ otherwise},
		\end{array}\right.
	\end{align}
	where we can identify $c_{0,\ldots,0,N_i=2,0,\ldots,0}=E(i,i)/E$ and $c_{0,\ldots,0,N_i=1,0,\ldots,0,N_j=1,0,\ldots,0}=(E(i,j)+E(j,i))/E$.
	To estimate the value $\overline f$ of a function $f_{N_0,\ldots,N_K}$, we can recast the standard sampling formula as follows:
	\begin{align}
		\nonumber
		\overline{f}
		=&\sum_{\substack{N_0,\ldots,N_K:\\N_0+\cdots+N_K=N}}f_{N_0,\ldots,N_K}\,c_{N_0,\ldots,N_K}
		\\\nonumber
		=&\sum_{0\leq i\leq K}\frac{E(i,i)}{E}
		\overbrace{f_{0,\ldots,0,N_i=2,0,\ldots,0}}^{\stackrel{\mathrm{def}}{=}f(i,i)}
		\\\nonumber
		&+\sum_{0\leq i<j\leq K}\frac{E(i,j)+E(j,i)}{E}
		\underbrace{f_{0,\ldots,0,N_i=1,0,\ldots,0,N_j=1,0,\ldots,0}}_{\stackrel{\mathrm{def}}{=}f(i,j)=f(j,i)}
		\\
		=&\frac{1}{E}\sum_{i,j=0}^K f(i,j)E(i,j).
	\end{align}
	See the Supplemental Material to Ref. \cite{SCEMRKNLGVAW17} for the generalization to $N>2$.

	In order to apply the multinomial framework described in the main text, one has to satisfy the premise that the coincidence statistics is symmetric, $E(k_1,k_2)=E(k_2,k_1)$.
	However, in reality, this is only true to a limited extent since the beam splitters in the multiplexing might not be perfectly balanced, and the detectors after the multiplexing might have slightly different responses.
	In Ref. \cite{SCEMRKNLGVAW17}, we provided a rough systematic error estimate to account for such imperfections, which is further refined in the following.

	The premises mentioned above state that the coincidences are symmetric.
	The actual measurements $E(k_1,k_2)$ naturally exhibit a certain amount of asymmetry;
	if not, no systematic error needs to be assigned.
	We can decompose the coincidences as follows:
	\begin{align}
		\label{eq:CoincidenceDecomposition}
		E(k_1,k_2)=
		\overbrace{\frac{E(k_1,k_2)+E(k_2,k_1)}{2}}^{\text{(symmetric part)}}
		+\overbrace{\frac{E(k_1,k_2)-E(k_2,k_1)}{2}}^{\text{(asymmetric part)}}.
	\end{align}
	Furthermore, assume we estimate a function $f(k_1,k_2)$ to obtain the mean $\overline{f}=\sum_{k_1,k_2} f(k_1,k_2)E(k_1,k_2)/E$.
	Inserting the above decomposition and denoting with $\overline{f}^\mathrm{(sym)}$ the value obtained from the symmetric part in Eq. \eqref{eq:CoincidenceDecomposition}, we apply the triangle inequality and find
	\begin{align}
		\label{eq:SystematicError}
		\left|\overline f-\overline{f}^\mathrm{(sym)}\right|\leq \sum_{k_1,k_2}|f(k_1,k_2)|\left|\frac{E(k_1,k_2)-E(k_2,k_1)}{2E}\right|=\epsilon_f,
	\end{align}
	which is the systematic error resulting from the asymmetry in the measured data.

	As we use the typical quadratic error propagation---rather than the linear form used for the above derivation---, we replace the right-hand-side expression in Eq. \eqref{eq:SystematicError} with $\epsilon_f^2=\sum_{k_1,k_2}|f(k_1,k_2)|^2\left|[E(k_1,k_2)-E(k_2,k_1)]/[2E]\right|^2$.
	Recall that the general relation between linear and quadratic error expansion for a function $F(x_1,x_2,\ldots)$ is given by $\Delta^\mathrm{(lin)} F=\sum_j |\partial F/\partial x_j|\Delta x_j$ and $\Delta^\mathrm{(quad)} F=(\sum_j |\partial F/\partial x_j|^2[\Delta x_j]^2)^{1/2}$.
	In addition, let us remind ourselves that the random error reads $\sigma_f=[(\overline{f^2}-\overline{f}^2)/(E-1)]^{1/2}$, which is combined with the systematic error to give the overall uncertainty, $\Delta f=\sqrt{\epsilon_f^2+\sigma_f^2}$.

\section{Sub-multinomial light}\label{sec:SubMultinomial}

	We assess our results with our previously derived nonclassicality criteria \cite{SCEMRKNLGVAW17,SECMRKNLGWAV17}.
	Let us briefly recapitulate this approach and its implementation for a self-consistent reading.
	The previously devised method is based on the observation that a correlation matrix, $M$, is positive semidefinite for classical light, i.e., $M=(M_{i,j})_{i,j=0,\ldots, K}\geq0$, where
	\begin{align}
	\begin{aligned}
		M_{i,j}
		=&N\overline{N_i(N_j+\delta_{i,j})}-(N-1)\overline{N_i}\,\overline{N_j}
		\\
		=&N^2(N-1)\left(
			\langle{:}\hat\pi_i\hat\pi_j{:}\rangle-\langle{:}\hat\pi_i{:}\rangle\langle{:}\hat\pi_j{:}\rangle
		\right),
	\end{aligned}
	\end{align}
	with $\delta$ denoting the Kronecker symbol.
	It was shown that the required first- and second-order moments can be obtained from coincidences as \cite{SCEMRKNLGVAW17}
	\begin{align}
		\overline{N_i}
		=&N\langle{:}\hat\pi_i{:}\rangle
		=\sum_{k_1,k_2}\left(\delta_{k_1,i}+\delta_{k_2,i}\right)\frac{E(k_1,k_2)}{E},
		\\\nonumber
		\overline{N_i(N_j+\delta_{i,j})}=&N(N-1)\langle{:}\hat\pi_i\hat\pi_j{:}\rangle
		\\
		=&\sum_{k_1,k_2}\left(\delta_{k_1,i}\delta_{k_2,j}+\delta_{k_1,j}\delta_{k_2,i}\right)\frac{E(k_1,k_2)}{E}.
	\end{align}
	Finally, the minimal eigenvalue $\mu_{\min}$ of the correspondingly reconstructed matrix $M$ is computed to probe for positive semidefiniteness.

	The heralding with a TES enables us to generate higher-order photon-number states.
	In Table \ref{tab:SubMultinomial}, we listed the nonclassicality in terms of the criteria $\mu_{\min}<0$ for data with a coherent amplitude zero.
	The observed nonclassicality in Table \ref{tab:SubMultinomial} for heralding bins larger than two is no longer significant within a three-standard-deviation error margin as the total number of events $E$ is too small in those cases.
	For this reason, we restrict our considerations to heralding bins $k_h\in\{0,1,2\}$.

\begin{table}[t]
	\caption{
		For the available heralding bins, $k_h$, the minimal eigenvalues $\mu_{\min}$ of the matrix $M$ are shown.
		Significant negativities defines the notion of nonclassical sub-multinomial light \cite{SCEMRKNLGVAW17,SECMRKNLGWAV17}.
	}\label{tab:SubMultinomial}
	\begin{tabular}{p{.2\columnwidth} p{.4\columnwidth} p{.3\columnwidth} }
		\hline\hline
		$k_h$ && sub-multinomial\\
		\hline
		0 && $-0.000\,0 \pm 0.000\,9$ \\
		1 && $-0.15\phantom{0\,0} \pm 0.03$ \\
		2 && $-0.10\phantom{0\,0} \pm 0.03$ \\
		3 && $-0.17\phantom{0\,0} \pm 0.07$ \\
		4 && $-0.3\phantom{00\,0} \pm 0.2$ \\
		\hline\hline
	\end{tabular}
\end{table}

\section{Local oscillator amplitudes}\label{sec:LocalOsci}

	From the first derivative of Eq. \eqref{eq:Photoelectric} for the photoelectric model, we can infer the dimensionless and displaced intensity,
	\begin{align}
		I(\beta)=&\left.\frac{\partial G_z(\beta)}{\partial z}\right|_{z=1}
		=|t|^2\eta\langle{:}\hat n\left(\frac{r}{t}\beta\right){:}\rangle_{\hat\rho}
		\\\nonumber
		=&\sum_{\substack{N_0,N_1,\ldots\geq0:\\N_0+N_1+\cdots=N}}(0N_0+1N_1+\cdots)c_{N_0,\ldots,N_K}(\beta),
	\end{align}
	where the latter expression results from the definition $G_z(\beta)=\sum_{N_0,N_1,\ldots}z^{0N_0+1N_1+\cdots}c_{N_0,\ldots,N_K}(\beta)$.
	Most importantly, in the case that the SI is vacuum, we find $\langle{:}\hat n(\gamma){:}\rangle_{|0\rangle\langle 0|}=|\gamma|^2$.
	We abstract this observation and define for the general detection scenario a detector-independent amplitude $|\beta^\mathrm{(DI)}|=\sqrt{I(\beta)}=\sqrt{\partial G_z(\beta)/\partial z|_{z=1}}$ for general POVMs and the SI $\hat\rho=|0\rangle\langle 0|$.
	This also explains the following Eq. \eqref{eq:EstimateAmplitude} as well as Eq. (7) in the main text.

\begin{figure}[t]
	\includegraphics[width=.9\columnwidth]{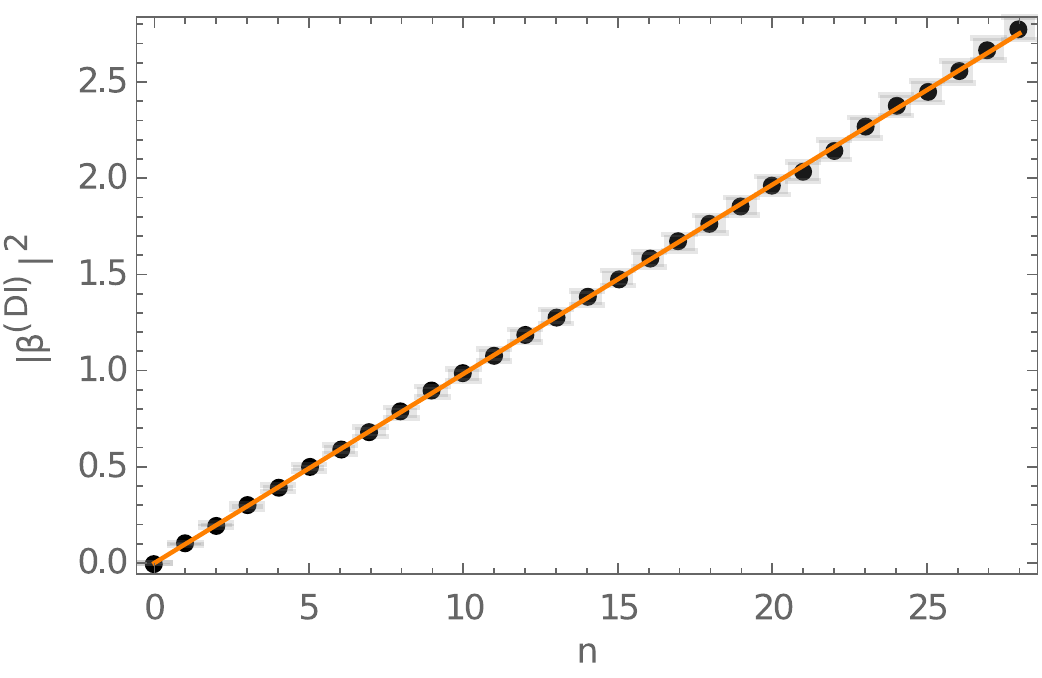}
	\caption{
		The estimated intensity [cf. Eq. \eqref{eq:EstimateAmplitude}] as a function of the used setting number $n$.
		The slope of the linear fit (orange line) is $0.1$.
		This confirms the intended difference of $1$ photon between two settings when correcting for the $90:10$ splitting that uses $10\%$ of the LO intensity.
	}\label{fig:Calibration}
\end{figure}

	We block the SI to infer the amplitude of the LO.
	The intensity, here represented through the dimensionless quantity
	\begin{align}
		\label{eq:EstimateAmplitude}
		\left|\beta^\mathrm{(DI)}\right|^2=\sum_{i=0}^K i\overline{N_i}\in[0,NK],
	\end{align}
	has been varied, realized via $n=0,\ldots,28$ power settings of the seed laser.
	These settings have been also applied when the SI is not blocked.
	The settings are chosen such that an equidistant intensity grid is generated.
	This is confirmed through the linear fit in Fig. \ref{fig:Calibration}.

\section{Additional analysis and results}\label{sec:Additional}

\subsection{Typical data and error estimates}\label{sec:SamplingError}

\begin{figure}[b]
	\includegraphics[width=.8\columnwidth]{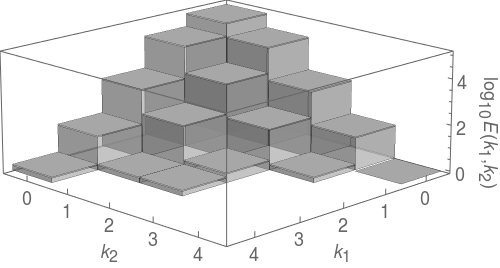}
	\caption{
		Raw coincidence data for $k_h=1$ and $\beta=0$.
		We recorded $E=184\,426$ events in $K+1=5$ bins.
	}\label{fig:RawData}
\end{figure}

	In order to further assess the impact of uncertainties, let us consider a typical example.
	In Fig. \ref{fig:RawData}, we depict a typical data set, there for a single photon (i.e., heralding to bin $k_h=1$) and a vanishing LO amplitude (i.e., the setting $n=0$).

\begin{figure}[t]
	\includegraphics[width=\columnwidth]{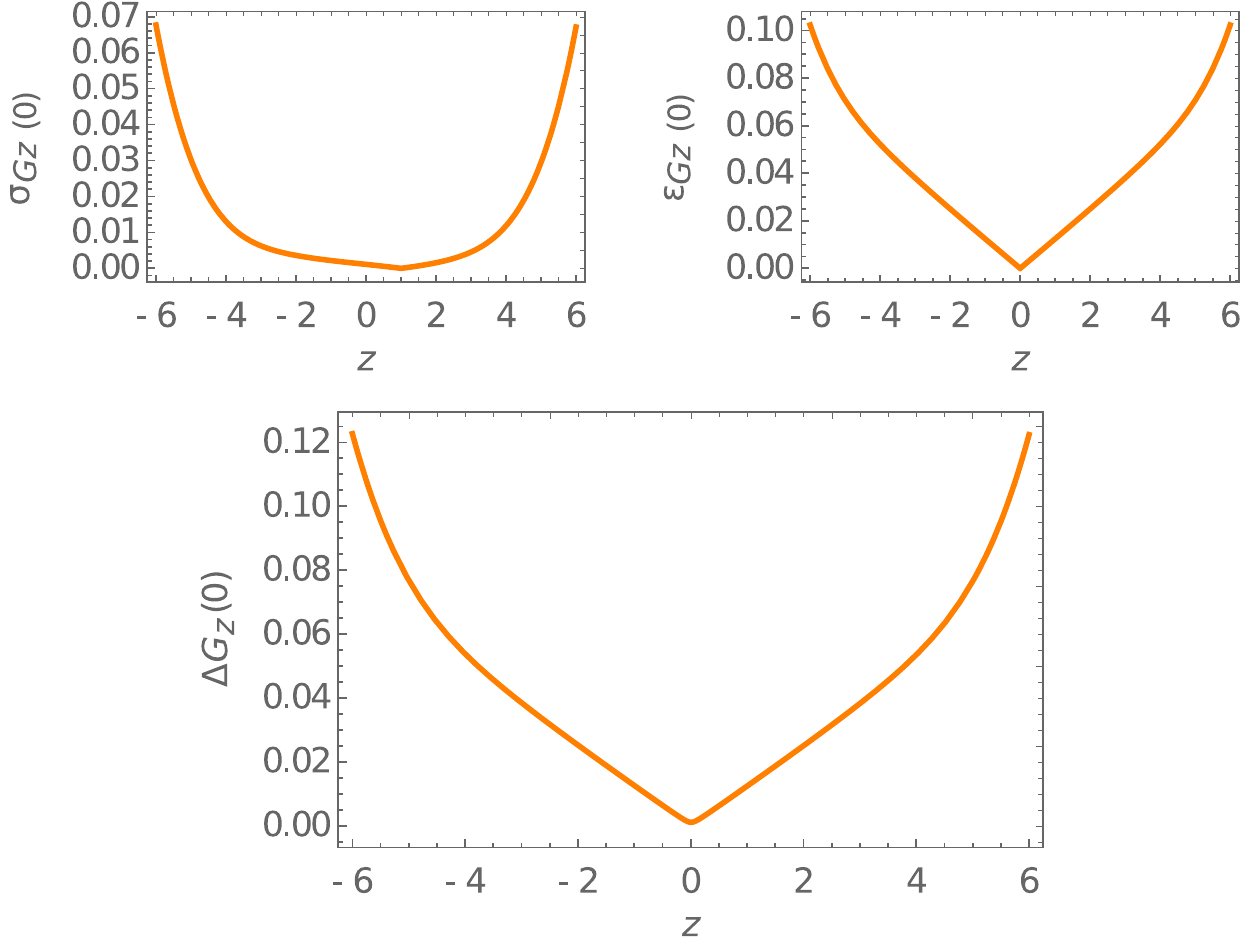}
	\caption{
		Error composition for the sampling of $G_z(\beta=0)$ as a function of $z$ for the data shown in Fig. \ref{fig:RawData}
		[top-left: random error, $\sigma_{G_z(0)}$; top-right: systematic error, $\epsilon_{G_z(0)}$; bottom: combination of both random and systematic uncertainties, $\Delta{G_z(0)}$].
	}\label{fig:Uncertainty}
\end{figure}

	The resulting (systematic and random) observational errors for the generating function are depicted in Fig. \ref{fig:Uncertainty}.
	The expected trend of monotonicity and the diverging behavior of the uncertainties as a function of $|z|$ are clearly visible.
	As $G_{z=1}(\beta)=1$ corresponds to the total probability, which is not subject to fluctuations, the random error vanishes for $z=1$ (Fig. \ref{fig:Uncertainty}, top-left panel).
	By construction, the systematic errors are symmetric with respect to $z=0$ (top-right plot in Fig. \ref{fig:Uncertainty}).
	See also the following discussion in Sec. \ref{sec:Quasiprobability}.

	For probing the nonclassicality through the generating function, obtained as $g_{z_0,\ldots,z_K}(\beta)=\sum_{k_1,k_2} z_{k_1}z_{k_2} E(k_1,k_2)/E$, we compute the eigenvector $Z=[z_0,\ldots,z_K]$ to the normalized coincidence matrix $[E(k_1,k_2)/E]_{k_1,k_2=0,\ldots,K}$ that corresponds to the minimal eigenvalue.
	For the example under study, we get $Z=[0.228, -0.973, -0.033, -0.001, -0.000]$.
	With this information, we can now estimate the general generating function and get $g_{z_0,\ldots,z_K}(\beta)=-0.026\pm0.003$ as the optimal negativity, here for $\beta=0$.
	The minimum over all measured LO amplitudes then yields $g_{\min}$, cf. Eq. (6) in the main text.

\subsection{Optimal quasiprobability distribution and error estimates}\label{sec:Quasiprobability}

\begin{figure}[t]
	\includegraphics[width=\columnwidth]{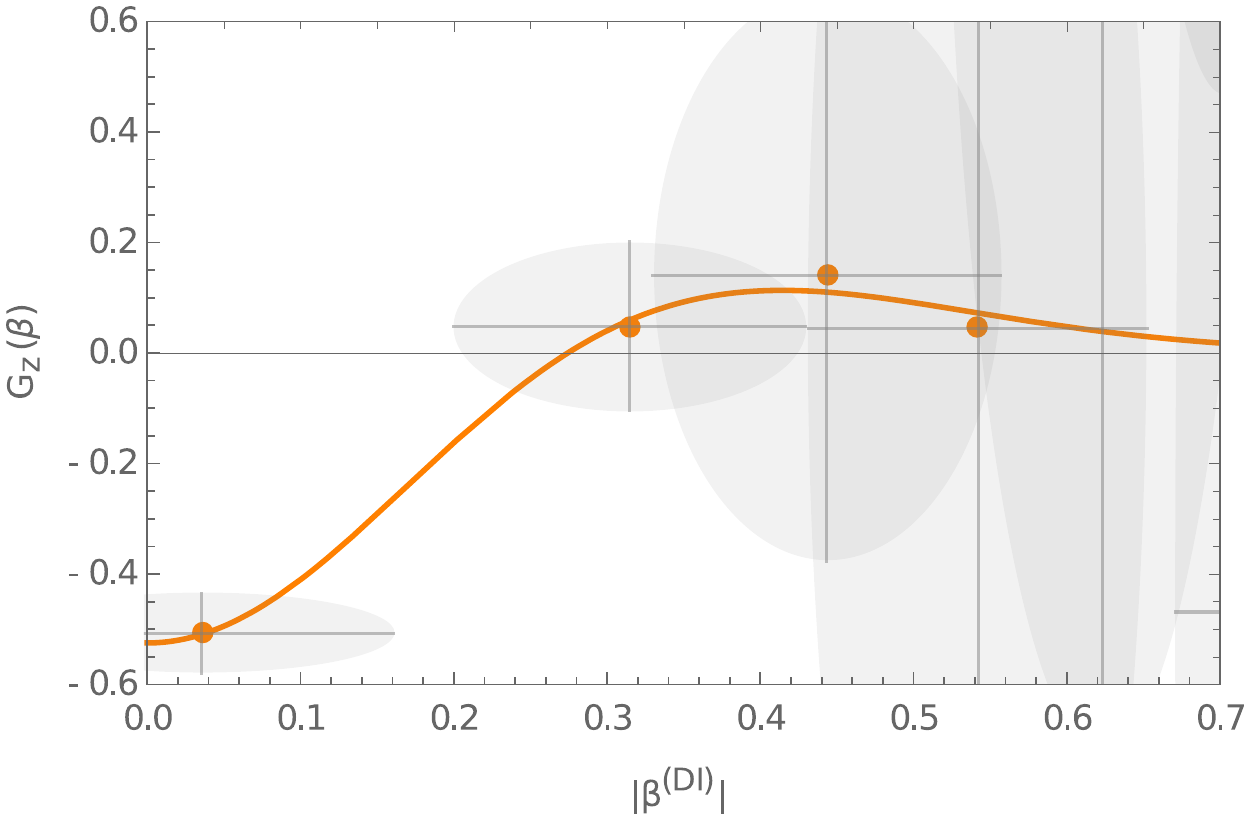}
	\caption{
		Radial component of the phase-space distribution of the heralded single-photon state ($k_h=1$) for the optimal value $z=-4.85$.
		For almost all data points with $|\beta^\mathrm{(DI)}|>\sqrt{0.5}$, the error estimates exceeds the plot range while allowing for consistency with expected mean values close to zero, cf. the discussion in Sec. \ref{sec:Quasiprobability}.
	}\label{fig:Quasiprobability}
\end{figure}

	We found that the parameter $z=-4.85$ is optimal in the sense that $G_z(\beta)$ has the most statistically significant negativity at the origin (see Fig. \ref{fig:Quasiprobability}), i.e., $-G_z(0)/\Delta G_z(0)$ is maximized.
	In the plotted scenario, the error estimates $\Delta G_z(\beta)$ are rapidly increasing for increasing $|\beta|$, which we discuss in the following based on the sampling formula
	\begin{align}
		G_z(\beta)=\sum_{N_0+\cdots+N_K=N} z^{0N_0+\cdots+KN_K} c_{N_0,\ldots,N_K}(\beta).
	\end{align}
	For increasing LO amplitudes, the components of $c_{N_0,\ldots,N_K}(\beta)$ that relate to a higher power of $z$ have a higher contribution to the estimate of this function.
	Similarly, $|z|> 1$ also leads to a most relevant term that corresponds to a higher power of $z$.
	Recall that the exponent $0N_0+\cdots+KN_K$ relates to the overall intensity [cf. Eq. \eqref{eq:EstimateAmplitude}].
	Consequently, both a large LO amplitude and $|z|> 1$ result in the fact that the contribution for $z^p$ for larger $p$ becomes the most relevant one.
	Standard error propagation then implies a relative error scaled by the large factor $p$, for increasing LO amplitudes and increasing $|z|$ values, which also explains why the increase of observational uncertainties in those scenarios is expected.

\section{Quantum state discrimination}\label{sec:Discrimination}

	As our distributions can, in principle, take arbitrary forms for arbitrary detectors, let us formulate the statistical model to discriminate states based on the reconstructed phase-space functions alone.
	The probability that two distributions, described through multivariate random variables $X$ and $X'$, are indistinguishable within a $\delta$-uncertainty can be expressed as
	\begin{align}
		\label{eq:Discrimination}
		\mathrm{Prob}(|X-X'|\leq \delta)=\int\limits_{-\delta}^{+\delta} du \int\limits_{-\infty}^{+\infty} dz\,p(X=z)p(X'=z+u),
	\end{align}
	where $p(X)$ and $p(X')$ are the probability densities of the uncertainties for the two random variables.
	We identify $X=[G^{(k_h)}_z(\beta_0),\ldots,G^{(k_h)}_z(\beta_{28})]$ and $X'=[G^{(k'_h)}_z(\beta_n)]_{n=0,\ldots,28}$ for different heralded states and LO settings and use a Gaussian error model with a mean and variance that corresponds to the reconstructed distributions for each measured setting $n$.
	Consequently, we get from Eq. \eqref{eq:Discrimination} the following probability for the discrimination:
	\begin{align}
		\nonumber
		&\mathrm{Prob}\left(G^{(k_h)}_z\neq G^{(k'_h)}_z\right)=1-\mathrm{Prob}(\forall n:|X_n-X_n'|\leq \delta_n)
		\\
		=&1-\prod_{n}
		\underbrace{
			\int\limits_{-\delta_n}^{+\delta_n}du_n\,
			\frac{
				\exp\left[-\frac{(u_n-[\mu_n-\mu_n'])^2}{2\Delta_n^2}\right]
			}{\sqrt{2\pi\Delta_n^2}}
		}_{
			=\mathrm{Err}\left[\frac{|\mu_n-\mu'_n|}{\Delta_n}+3\right]
			-\mathrm{Err}\left[\frac{|\mu_n-\mu'_n|}{\Delta_n}-3\right]
		}
		,
	\end{align}
	where we set the vector $\delta$ to correspond to three combined standard deviations for each setting, $\delta_n=3\Delta_n$ and $\Delta_n=[\Delta G^{(k_h)}_z(\beta_n)^2+\Delta G^{(k_h)}_z(\beta_n)^2]^{1/2}$, and using the mean values $\mu_n=\overline{G^{(k_h)}_z(\beta_n)}$ and $\mu'_n=\overline{G^{(k'_h)}_z(\beta_n)}$.
	Note that $\mathrm{Err}[z]=\int^z d\xi e^{-\xi^2/2}/\sqrt{2\pi}$ denotes the error function.

	For instance, we find for our measured data that the likelihood to discriminate the phase-space distributions for $k_h$ from the one for $k'_h$ for $z=-1.5$ is given by the matrix
	\begin{align}
		\!\!\!\!\left[\!\mathrm{Prob}\!\!\left(\!\! G^{(k_h)}_z{\neq} G^{(k'_h)}_z \!\!\right)\!\!\right]_{\!\! k_h,k'_h{=}0,1,2}\!\!
		{=}\!\!\!\begin{bmatrix}
			7.5\% & 100.\% & 100.\%
			\\
			100.\% & 7.5\% & 98.9\%
			\\
			100.\% & 98.9\% & 7.5\%
		\end{bmatrix}\!\!\!,
	\end{align}
	where ``$100.\%$'' corresponds to a value which is $100\%$ within the used numerical precision.
	Note that the diagonal elements are nonzero as identical distributions could still represent different states when considering a finite error margin.

\section{Fit model from vacuum measurements}\label{sec:Gauge}

	From the measurement in which the SI is blocked (i.e., vacuum SI), we can extrapolate the general shape of the phase-space distribution for photon states, without relying on any particular detector model.
	This approach is also used to fit the reconstructed distributions for the heralding to $k_h$.

\begin{figure}[b]
	\includegraphics[width=.8\columnwidth]{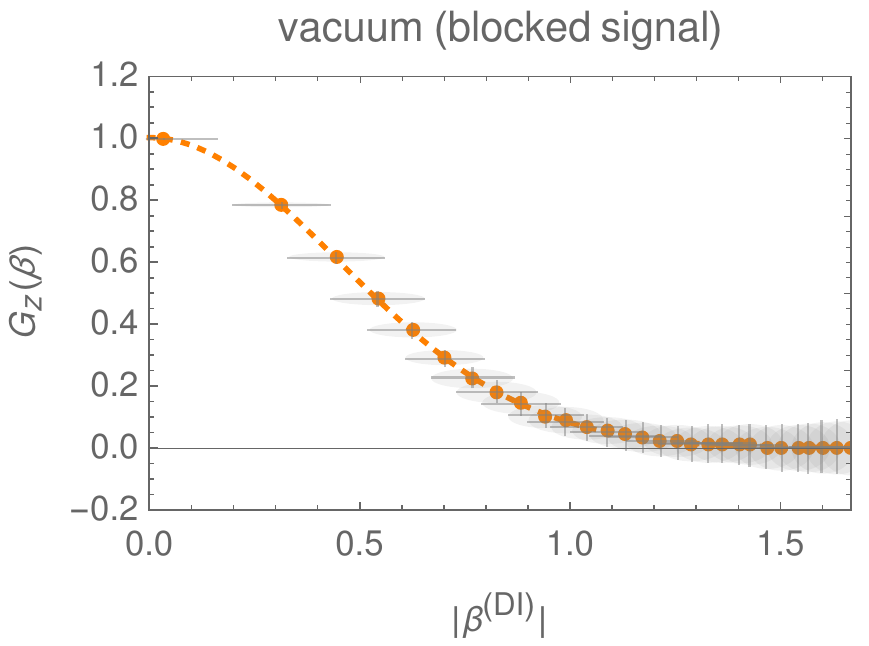}
	\caption{
		Phase-space distribution for vacuum, $G_z(\beta)=G_z^\mathrm{(vac)}(\beta)=G^{(0)}_z(\beta)$, with $z=-1.5$.
		The dashed line corresponds to a fit function $f_{0}\exp[-b|\beta^\mathrm{(DI)}|^2]$ for real-valued constants $f_0$ and $b$.
	}\label{fig:Vacuum}
\end{figure}

	Using the data where the signal is blocked, we find that a Gaussian distribution describes the reconstructed phase-space distribution for vacuum quite well; see Fig. \ref{fig:Vacuum}.
	This information can be used to predict the phase-space distributions of $m$-photon states as well.
	Because of Eq. \eqref{eq:GeneratingFunction2} and the known representation $P^{(m)}(\alpha)=\sum_{j=0}^m\binom{m}{j}j!^{-1}\partial_{\alpha}^j\partial_{\alpha^\ast}^jP^{(0)}(\alpha)$ \cite{VW06}, where $P^{(0)}(\alpha)$ describes the delta distribution centered at the origin, we find that the $m$-th photon state is given by
	\begin{align}
		\label{eq:FockStates}
		G^{(m)}_z(\beta)=\sum_{j=0}^m\binom{m}{j}\frac{1}{j!}\left[\frac{|t|^2}{|r|^2}\right]^j\partial_{\beta}^j\partial_{\beta^\ast}^jG^{(0)}_z(\beta),
	\end{align}
	where $G^{(0)}_z(\beta)$ is experimentally obtained by blocking the signal (Fig. \ref{fig:Vacuum}) and which is determined without relying on any detection models.
	For deriving Eq. \eqref{eq:FockStates}, note that the argument of the vacuum function implies that a partial integration of Eq. \eqref{eq:GeneratingFunction2} with derivatives of delta distributions results in $\left.\partial^j_\alpha\partial^{j}_{\alpha^\ast} f(\beta-t\alpha/r)\right|_{\alpha=0}=(-t/r)^j(-t^\ast/r^\ast)^{j}\partial^j_\beta\partial^{j}_{\beta^\ast} f(\beta)$.

	From Eq. \eqref{eq:FockStates} and the fit obtained from $G^{(0)}_z(\beta)$, we can therefore predict the phase-space distribution of an $m$-photon state.
	In our case, this means that $G^{(m)}_z(\beta)$ is a Gaussian function multiplied with a fixed $m$th-order polynomial in $|\beta|^2$.
	In this context, also recall the linear relation between the actual intensity (via the setting number $n$) and the detector-independent intensity in Fig. \ref{fig:Calibration}.
	In addition, it is known (see, e.g., Ref. \cite{SVA14}) that imperfect heralding for the kind of photon source used leads to additional noise contributions.
	For this reason, our fit for an heralding to the $k_h$th bin is described through $G^{(k_h)}_z(\beta)=\sum_{j=0}^{k_h} f_j|\beta^\mathrm{(DI)}|^{2j}\exp[-b|\beta^\mathrm{(DI)}|^2]$, which constitutes the generalized fit function used in the main text [Figs. 2(a)--(c)] and is determined from the vacuum measurements alone and without relying on any detection models.

	As a final remark, it is worth mentioning that the above treatment can be straightforwardly generalized to predict $G_z(\beta)$ for arbitrary states, resulting in Eq. (8) in the main text.
	This is based on the fact that the $P$ function of an arbitrary state can be written as a convolution, $P(\alpha)=\int d^2\alpha'\,P(\alpha-\alpha')P^{(0)}(\alpha')$, recalling that the vacuum state is described by a delta distribution, $P^{(0)}$.
	Thus, Eq. \eqref{eq:GeneratingFunction2} implies that $G_z$ of an arbitrary state, represented through the Glauber-Sudarshan distribution $P$, is predicted to resemble the convolution of the already measured vacuum state's $G_z^{(0)}$ and $P$.
	Even more generally, we can write
	\begin{align}
		g_{z_0,\ldots,z_K}(\beta)=\int d^2\alpha\, P(\alpha) g^\mathrm{(0)}_{z_0,\ldots,z_K}\left(\beta-\frac{t}{r}\alpha\right),
	\end{align}
	where $g^\mathrm{(0)}_{z_0,\ldots,z_K}(\beta)=\left[\sum_{k=0}^K z_kp_k\left(-r\beta/\sqrt N\right)\right]^N$, cf. Eq. \eqref{eq:GeneratingFunction}, to predict the phase-space distribution for a state, theoretically described through $P(\alpha)$, via the measured $g^\mathrm{(0)}_{z_0,\ldots,z_K}$.


\begin{thebibliography}{99}
\bibitem{commentsParameter}
		Note that in our case, the $z$ parameter is not limited to values that correspond to $-1\leq s\leq 1$ because of $\mathcal K$ being a finite set and, thus, guaranteeing convergence of $G_z$ for any $z$.
\end{thebibliography}
\end{document}